\documentclass[]{svjour2}

\usepackage{lmodern}
\usepackage{amsmath}
\usepackage{amssymb}
\usepackage{mathptmx}
\usepackage{graphicx}
\usepackage{bm}
\usepackage{comment}
\usepackage{rotating}
\usepackage[numbers]{natbib}

\bibpunct{}{}{,}{s}{}{,}

\newcommand{\ci}{i}
\newcommand{\cpi}{\pi}
\newcommand{\ce}{e}
\newcommand{\dd}{d}
\newcommand{\average}[1]{\left\langle#1\right\rangle}

\newcommand{\norm}[1]{\lvert#1\rvert}

\newcommand{\coloronline}{(color online) }

\makeatletter
\journalname{Journal of Low Temperature Physics}

\begin{document}

\setlength{\abovedisplayskip}{5pt}
\setlength{\belowdisplayskip}{5pt}

\title{Impurity effects in a vortex core in a chiral $p$-wave superconductor within the $t$-matrix approximation}


\author{Noriyuki Kurosawa$^1$ \and Nobuhiko Hayashi$^2$ \and Emiko Arahata$^3$ \and Yusuke Kato$^1$}
\institute{%
  1:Department of Basic Science, The University of Tokyo, Komaba, Meguro, Tokyo 153-8902, Japan
\\
  2:NanoSquare Research Center (N2RC), Osaka Prefecture University, 1-2 Gakuen-cho, Naka-ku, Sakai 599-8570, Japan\\
  3:Institute of Industrial Science, The University of Tokyo, Komaba, Meguro, Tokyo 153-8505, Japan
}

\date{\today}

\maketitle
\begin{abstract}
  We study the effects of non-magnetic impurity scattering on the Andreev bound states (ABS) in an isolated vortex in two-dimensional chiral $p$-wave superconductors numerically. We incorporate the impurity scattering effects into the quasiclassical Eilenberger formulation through the self-consistent $t$-matrix approximation. Within this scheme, we calculate the local density of states (LDOS) around two types of vortices: ``parallel'' (``anti-parallel'') vortex where the phase winding of the pair-potential coming from the vorticity and that coming from the chirality have the same (opposite) sign. 

When the scattering phase-shift $\delta_0$ of each impurity is small, we find that the impurities affect differently the spectra of quasiparticles localized around the two types of vortex in a way similar to that in the Born limit ($\delta_0\to 0$). For a larger $\delta_0(\lesssim \pi/2)$, ABS in the vortex is strongly suppressed by the impurities for both types of vortex. 
We find that there are some correlations between the suppression of ABS near vortex cores and the low energy density of states due to the impurity bands in the bulk.
\end{abstract}


\section{Introduction}
Chiral superconductors/superfluids are referred to as those in which time-reversal and two-dimensional parity symmetries as well as the gauge symmetry are broken spontaneously. 
%
%
%
Two-dimensional chiral $p$-wave superconductor/superfluid, which is believed to be realized in Sr$_\text{2}$RuO$_\text{4}$\cite{Maeno1994-Nature372-532, Mackenzie2003-RMP75-657, Sigrist2005-PTPS160-1, Maeno2012-JPSJ81-011009} and thin film superfluid ${}^\text{3}$He-A\cite{Vorontsov2003-PRB68-064508}, is a spin-triple superconductor/superfluid specified by $d$-vector ${\bm d}=\hat{z}(k_x +i k_y)/k_{\rm F}$ or ${\bm d}=\hat{z}(k_x -i k_y)/k_{\rm F}$, each state of which corresponds to an eigenstate of the internal angular momentum (, which we call chirality) of its Cooper pair. In the absence of external magnetic fields, these two states are degenerate and one of them is selected spontaneously as the thermodynamic state.

Under a symmetry-broken state, isolated vortex and antivortex in the chiral superconductor are physically inequivalent\cite{Heeb1999-PRB59-7076,Tokuyasu:1,Matsumoto:1,Kato2001-JPSJ70-3368}. One type of vortex has the vorticity same as that of the chirality of the Cooper pair (``parallel vortex'') while the other type of vortex has the vorticity opposite to the chirality (``anti-parallel vortex''). The total angular momentum of the system $l_z$ of parallel vortex is $\pm 2\hbar$ and that of the anti-parallel vortex is $0$. 

Several authors have considered differences between the two types of vortices such as core energies\cite{Heeb1999-PRB59-7076,Tokuyasu:1} and lower critical field\cite{Heeb1999-PRB59-7076,Tokuyasu:1}. The authors of refs.~\cite{Matsumoto1999-JPSJ68-724,Kato2000-JPSJ69-3378,Kato2002-JPSJ71-1721,Hayashi2005-JLowTempPhys139-79, Tanuma2009-PRL102-117003} argued that in the anti-parallel vortices the effect of the non-magnetic impurity (impurities) on the Andreev bound states (ABS) in the core\cite{Caroli1964-PhysLett9-307,Stone1996-PRB54-13222} is considerably suppressed compared to the parallel vortices. This effect has also been reported in a similar system\cite{Yokoyama2008-PRL100-177002} and some authors have ascribed this robustness to ``odd-frequency pairing''\cite{Tanuma2009-PRL102-117003,Tanaka2012-JPSJ81-011013}. The vortex-type dependent impurity effects are important in the sense that these effects imply vortex-type-dependent flux-flow conductivity and Hall conductivity in chiral superconductors.

However, these studies (except ref.~\cite{Matsumoto1999-JPSJ68-724}, where the effect of a single impurity located at the vortex center was considered) have been done only in the Born limit, which corresponds to the situation where a lot of weak scatterers exist randomly, and the scattering phase-shift of a single impurity potential $\delta_0$ is extremely small ($\delta_0\rightarrow 0$) in the $t$-matrix formulation. In general, the impurity effect on unconventional superconductors depends crucially on $\delta_0$. For example, experimental results on thermal conductivity in heavy fermion superconductors were explained by the theories taking account of impurity effects due to unitary scatterers ($\delta_0\to\cpi/2$)\cite{SchmittRink,Hirshfeld}. 
For another example, Hayashi \textit{et al}.\ have reported that the core-shrinkage effect\cite{Kramer1974-ZPhys269-59} of vortex in the two-dimensional $s$-wave and chiral $p$-wave superconductors are different between the Born and unitary limits\cite{Hayashi2013-PhysicaC484-69,Hayashi2013-PhysicaC-InPress}. Because $\delta_0$ depends on the superconducting material and species of the impurities (for example, see ref.~\cite{Hill2004-PRL92-027001}), it is therefore not sufficient for adapting these theories to the real materials.
In addition, some authors\cite{Sauls2009-NewJPhys11-075008} have reported a result different from those obtained in refs.~\cite{Kato2000-JPSJ69-3378,Kato2002-JPSJ71-1721,Hayashi2005-JLowTempPhys139-79, Tanuma2009-PRL102-117003} in the Born limit. Thus the vortex-type dependent impurity effects in the Born limit itself is still an issue. 

To clarify the impurity effects on the ABS in vortices (in this paper, we call those states as vortex-ABS), we study both Born and unitary limits and the intermediate regime between them in a fully self-consistent way. We use Eilenberger's quasiclassical theory with the $t$-matrix formulation\cite{Eilenberger1968-ZPhys214-195,Thuneberg1984-PRB29-3913}. 
%
%
%
%
%
%
%
\section{Model and Method}
We consider two-dimensional spin-triplet chiral $p$-wave superconductors with isotropic circular Fermi surface in the type II limit,
i.e.\ the ratio of the magnetic penetration depth to the coherence length is taken to be infinity.
%

In the quasiclassical theory of superconductivity\cite{Eilenberger1968-ZPhys214-195}, the electronic structure of quasiparticles is described in terms of the quasiclassical Green's function
\begin{align}
  \check g(\ci\omega_n, \bm{r}, \bm{k}) &=
  \begin{pmatrix} g & f \\ -f^\dag & -g \end{pmatrix},
\label{eq: quasiclassical}
\end{align}
which is defined by the Gor'kov Green's function integrated over the magnitude of quasiparticle energy. 
The quasiclassical Green's function (\ref{eq: quasiclassical})
satisfies the Eilenberger equation\cite{Eilenberger1968-ZPhys214-195}
\begin{align}
  -\ci\hbar\bm{v}_{\text{F}}\cdot\nabla\check g &= \left[\ci\hbar\omega_n\check\tau_3-\check\Delta-\check\Sigma,\;\check g\right],
\label{eq: Eilenberger}
\end{align}
and the normalization condition $\check g^2 = -\cpi^2\check\tau_0$. 
Here $\check \tau_i$ $(i=0,1,2,3)$ denote the Pauli matrices in the particle-hole space. The symbol $\omega_n = (2n+1)\cpi k_{\text{B}}T/\hbar$ denotes the Matsubara frequency and $\bm{v}_{\text{F}}$, $\check\Sigma$ and $\check \Delta$ are, respectively, the Fermi velocity, the impurity self-energy and the pair-potential.

Within the $t$-matrix approximation\cite{Thuneberg1984-PRB29-3913}, the character of the impurities is parametrized by the scattering rate $\Gamma_{\text{n}}=\hbar(2\tau_\text{n})^{-1}$ or the relaxation time $\tau_{\text{n}}$ in the normal state and the phase-shift of a single impurity $\delta_0$. 
The impurity self-energy $\check\Sigma$
is expressed in terms of $\check g$, $\Gamma_{\rm n}$ and $\delta_0$ as \cite{Hayashi2005-JLowTempPhys139-79}
\begin{align}
  \check\Sigma(\ci\omega_n, \bm{r})
&= \frac{\cpi^{-1}\Gamma_\text{n}\average{\check g}}{\cos^2\delta_0-\cpi^{-2}\sin^2\delta_0(\average{g}^2-\average{f}\average{f^\dag})}.
\label{eq: t-matrix}
\end{align}
The notation $\average{A}$ denotes $A$ averaged over the Fermi surface, and in this case, it can be expressed as $\average{A} = \int_0^{2\cpi} \dd\alpha A(\bm{k})/(2\cpi)$ where $\bm{k}/k_{\text{F}} = (\cos\alpha, \sin\alpha)$.

The pair-potential has the matrix form of 
\begin{align}
\check \Delta(\bm{r}, \bm{k}) &= \begin{pmatrix} 0 & \Delta(\bm{r}, \bm{k}) \\ -\Delta^*(\bm{r}, \bm{k}) & 0\end{pmatrix},
\end{align}
where $\Delta(\bm{r},\bm{k})$ satisfies the gap equation\cite{Kato2002-JPSJ71-1721,Hayashi2005-JLowTempPhys139-79, Sauls2009-NewJPhys11-075008}
\begin{align}
  \Delta(\bm{r},\bm{k}) &= \lambda k_{\text{B}}T\sum_{n,\; \norm{\omega_n}\le\omega_{\text{c}}} \average{2\cos(\alpha-\alpha')f(\ci\omega_n,\bm{r},\alpha')}_{\alpha'}.
\label{eq: gap}
\end{align}
Here $\lambda$ is the coupling constant that satisfies\cite{Hayashi2005-JLowTempPhys139-79}
\begin{align}
  \frac{1}{\lambda} &= \ln\frac{T}{T_{\text{c0}}}+\sum_{n=0,\;\omega_n\le\omega_{\text{c}}}\frac{1}{n+1/2}
  .
\end{align}

The symbol $\omega_\text{c}$ denotes a cut-off frequency and we set $\hbar\omega_\text{c}=10\Delta_0$ as the same value in the earlier studies\cite{Hayashi2005-JLowTempPhys139-79, Tanuma2009-PRL102-117003}, where $\Delta_0$ is the modulus of the pair-potential at zero temperature in the bulk without the impurities.
When there exist impurities, the critical temperature $T_{\text{c}}$ obeys the Abrikosov-Gor'kov law for anisotropic superconductors\cite{Larkin1965-JETPL2-130}. Note that $T_{\text{c}}$ does not depend on $\delta_0$\cite{Mineev}.

In the absence of external magnetic fields, the chiral $p$-wave superconductors have two-fold degenerate thermodynamic states with the pair-potential $\Delta(\bm{k})\propto {\rm exp}(\pm {\rm i}\alpha)$; each state has the Cooper pairs with internal angular momentum $\pm \hbar$. In the chiral $p$-wave states with a single vortex with a positive vorticity at $r=0$, the pair-potential $\Delta(\bm{r},\bm{k})$ has the asymptotic form $\Delta_{\text{b}}\ce^{\ci(\phi\pm \alpha)}$ with $\bm{r}/r=(\cos\phi, \sin\phi)$ and the modulus in the bulk $\Delta_{\text{b}}$ far away from the vortex center in the presence of impurities at finite temperatures. In the intermediate regime with finite $\bm{r}$ around the single vortex, both Cooper pairs with $\pm \hbar$ coexist. Taking account of axial symmetry around $r=0$, we can write the pair-potentials \cite{Hayashi2005-JLowTempPhys139-79} generally in the forms 
\begin{align}
  \Delta^\text{(p)}(\bm{r},\bm{k}) &= \Delta_+^\text{(p)}(r)\ce^{\ci(\phi+\alpha)} + \Delta_-^\text{(p)}(r)\ce^{\ci(3\phi-\alpha)}
  & (\text{parallel vortex})
  \label{eq:phase-of-p-vortex}
  \\
  \Delta^\text{(a)}(\bm{r},\bm{k}) &= \Delta_+^\text{(a)}(r)\ce^{\ci(\phi-\alpha)} + \Delta_-^\text{(a)}(r)\ce^{\ci(-\phi+\alpha)}
  & (\text{anti-parallel vortex})
  \label{eq:phase-of-a-vortex}
\end{align}
respectively.
The subscripts $+$ and $-$ describe dominant and induced components of pair-potential; the latter component vanishes far away from the vortex center (i.e., $\Delta_-^\text{(p)}(r\rightarrow \infty)=0$, $\Delta_-^\text{(a)}(r\rightarrow \infty)=0$).

We numerically calculate the quasiclassical Green's functions $\check{g}$ around the isolated vortex in a self-consistent way through successive iterations of the Eilenberger equation \eqref{eq: Eilenberger}
, the Dyson equation \eqref{eq: t-matrix} and the gap equation \eqref{eq: gap}
for Matsubara frequency $\ci\omega_n$ and real-frequency $\epsilon$.
We solve eq.~\eqref{eq: Eilenberger} with use of the Riccati-parametrization\cite{Nagato1993-JLowTempPhys93-33, Schopohl1995-PRB52-490}. Equation (\ref{eq: Eilenberger}) is solved on the line (so called ``quasiclassical trajectory'') with a constant $b=\bm{r}\cdot(\hat{\bm{z}}\times\bm{k})$.
Note that $b$ can be regarded as the impact parameter, which is the quasiparticle angular momentum divided by the Fermi momentum. 
We adopt the classical fourth-order Runge-Kutta method to solve \eqref{eq: Eilenberger},
and for the initial value at $r=\pm 100\cpi\xi_0$ we use the bulk solution,
here we use symbol $\xi_0=2 E_\text{F}(\cpi k_\text{F}\Delta_0)^{-1}$ for the coherence length at zero temperature without impurities.

\section{Results and Discussion}
\begin{figure}
  \centering
  \includegraphics[bb=0 0 372 152, width=0.9\columnwidth, trim=5 10 0 20]{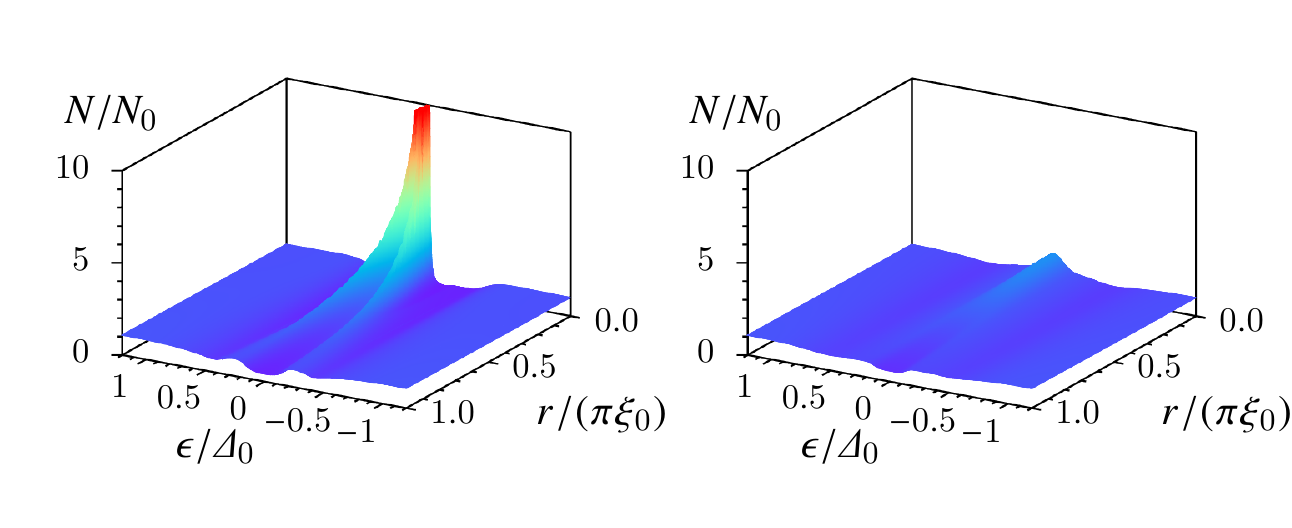}
  \caption{\coloronline The LDOS in the vortex in the Born limit ($\delta_0=0$) at $T/T_{\text{c0}}=0.3$ and $\Gamma_\text{n}/\Delta_0=0.3$. left: $l_z=0$, right: $l_z=2\hbar$}
  \label{FIG:LDOS-d00t01g03}
\end{figure}

Figures \ref{FIG:LDOS-d00t01g03} and \ref{FIG:LDOS-d10t01g03} show LDOS in the vortex at $T/T_{\text{c0}}=0.1$, $\Gamma_\text{n}/\Delta_0=0.3$ in the Born limit ($\delta_0\rightarrow 0$) and the unitary limit ($\delta_0\rightarrow \cpi/2$), respectively. In the Born limit, there is a sharp peak at zero energy at the center of anti-parallel vortex $(l_z =0)$ but the peak is suppressed for the parallel vortex $(l_z = 2\hbar)$. This implies that the low energy bound states (vortex-ABS) in the anti-parallel vortex
are more robust against the impurities than those in the parallel vortex. 
This behavior is consistent with earlier results\cite{Kato2000-JPSJ69-3378, Kato2002-JPSJ71-1721,Hayashi2005-JLowTempPhys139-79,Tanuma2009-PRL102-117003}.
We note that the level broadening with use of analytical expression for quasi-classical Green function obtained in ref.~\cite{Kato2002-JPSJ71-1721} within non-self-consistent Born approximation support our results and those of \cite{Tanuma2009-PRL102-117003}.
Contrarily to the above observations, it was reported in ref.~\cite{Sauls2009-NewJPhys11-075008} that the level broadening of core states due to impurities in the Born approximation does not depend on the type of vortices so much. As pointed out by one of the author in ref.~\cite{Sauls2009-NewJPhys11-075008}, this discrepancy may be attributed to difference in the presence or absence of constraint on axial symmetry\cite{SaulsPrivate}; in this study and ref.~\cite{Tanuma2009-PRL102-117003}, axial symmetry is imposed and the phase of the order parameter is set as \eqref{eq:phase-of-p-vortex} and \eqref{eq:phase-of-a-vortex} while in ref.~\cite{Sauls2009-NewJPhys11-075008}, the calculation was performed so that the spontaneous breaking of axial symmetry is allowed. The resultant solution
 in ref.~\cite{Sauls2009-NewJPhys11-075008}, however, preserves the axial symmetry. Thus, the constraint on axial symmetry is unlikely to affect the final result; the reason for the discrepancy of the results on the level-broadening in the Born approximation between the present study and ref.~\cite{Sauls2009-NewJPhys11-075008} is still unclear. 

%
%

In the unitary limit, on the other hand, the zero energy peak is considerably suppressed in the anti-parallel vortex
as well as the parallel vortex.
%
We quantify the effects of the impurities on the vortex-ABS by the peak value of LDOS at the vortex center $N(r=0,\epsilon=0)/N_0$, which we show in fig.~\ref{FIG:LDOS-peak}. The peak is suppressed more considerably when $\delta_0$ increases from $0$ (the Born limit) to $\cpi/2$ (the unitary limit).
%
For fig.~\ref{FIG:LDOS-peak}-(d), the peak of anti-parallel vortices ($l_z=0$) is suppressed even in the Born limit. We will discuss this behavior later. 

\begin{figure}
  \centering
  \includegraphics[bb=0 0 372 152, width=0.9\columnwidth, trim=5 10 0 20]{n-d00t03g03.pdf}
  \caption{\coloronline The LDOS in the vortex in the unitary limit ($\delta_0=\cpi/2$) at $T/T_{\text{c0}}=0.3$ and $\Gamma_\text{n}/\Delta_0=0.3$. left: $l_z=0$, right: $l_z=2\hbar$}
  \label{FIG:LDOS-d10t01g03}
\end{figure}

\begin{figure}
  \centering
  \includegraphics[bb=0 0 381 272, width=0.75\columnwidth]{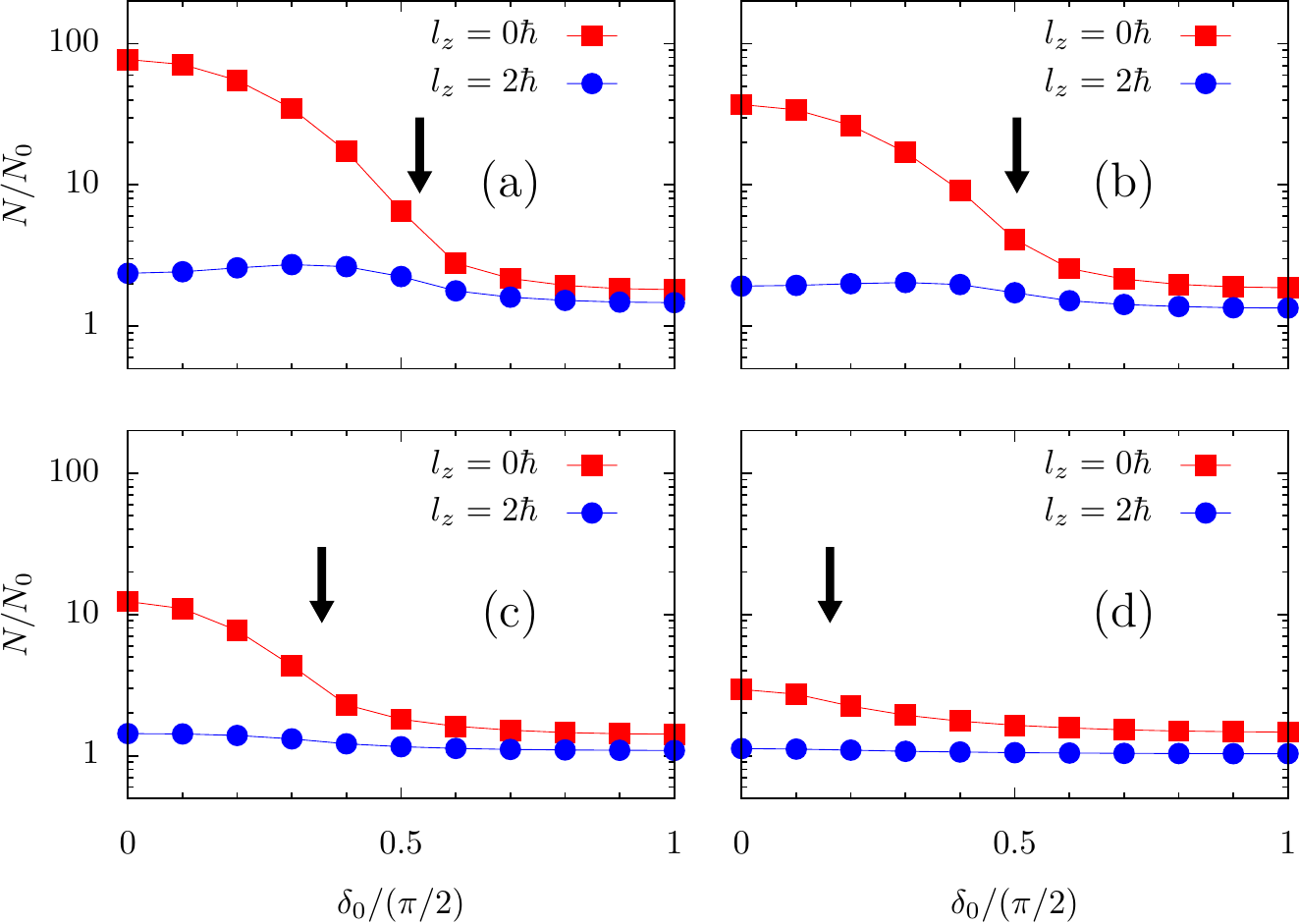}
  \caption{\coloronline The dependence of the peak value of zero energy LDOS at the center of the vortex upon the scattering phase-shift $\delta_0$. The temperature $T$ and scattering rate $\Gamma_\text{n}$ are $T/T_{\text{c0}}=0.1$ and $\Gamma_\text{n}/\Delta_0=0.3$ (a), $T/T_{\text{c0}}=0.3$ and $\Gamma_\text{n}/\Delta_0=0.3$ (b), $T/T_{\text{c0}}=0.1$ and $\Gamma_\text{n}/\Delta_0=0.4$ (c) and $T/T_{\text{c0}}=0.3$ and $\Gamma_\text{n}/\Delta_0=0.4$ (d). The arrows indicate $\delta_0$ where the energy gap on the Fermi level in the bulk disappears. The critical values are numerically calculated using \eqref{EQ:gap-closing-phase-shift}.}
    \label{FIG:LDOS-peak}
\end{figure}

  %

\begin{figure}
  \centering
  \includegraphics[bb=0 0 239 135, width=0.55\columnwidth]{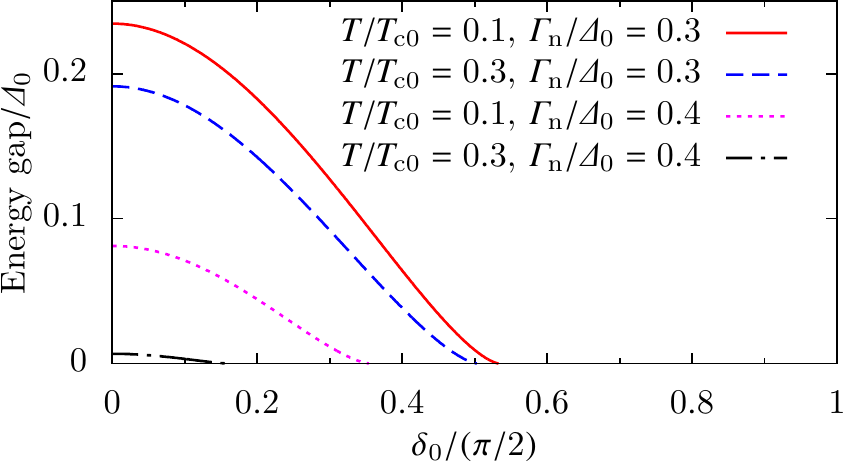}
  \caption{\coloronline The minimum excitation energy in the bulk.}
  \label{FIG:EnergyGap-size}
\end{figure}

In the following, we discuss the results shown in figs.~1-3, considering the impurity effects on the quasiparticle density of states (DOS) in chiral $p$-wave superconductors in the bulk.
%
%
%
The modulus of the pair-potential in the bulk is suppressed regardless of the type of the impurities when $T$ is high and $\Gamma_{\text{n}}$ is large. The reduction of $\Delta_{\text{b}}$ makes the vortex-ABS more extended spatially and lower the peak of LDOS near the core, regardless of the value of $\delta_0$.


Even when $\Delta_{\text{b}}$ is not so small (i.e., $T$ is sufficiently low and $\Gamma_{\text{n}}$ is sufficiently small), 
there exist quasiparticles with energy smaller than $\Delta_{\text{b}}$, which stem from the impurity bands. 
Following the standard calculation of the impurity effects in the spatially uniform unconventional superconductors \cite{Mineev}, we can obtain the gap edge.
When $\delta_0$ is increasing toward $\cpi/2$, the minimum energy for excitation decreases as shown in fig.~\ref{FIG:EnergyGap-size}, and it becomes zero (i.e., there are finite density of states at the Fermi level and a gapless superconductivity comes out\cite{Maki1999-EPL45-263}) when $\delta_0$ exceeds a critical value $\delta_{\rm c}$. 
The value of $\delta_{\rm c}$ is given as the solution of the equation
\begin{align}
  \delta_{\text{c}} &=\arccos\sqrt{\Gamma_{\text{n}}/\left\lvert\Delta_{\text{b}}(\Gamma_{\text{n}};\delta_{\text{c}})\right\rvert} \label{EQ:gap-closing-phase-shift}
\end{align}
under the assumption that $\Delta_{\text{b}}$ is monotonically decreasing of $\delta_0$. 
%
%
At the energy where the impurity band has finite density of states, there is a resonance between the localized wave functions near vortex cores and wave functions that extend spatially outside vortex and thus we can naturally understand the reason why the spectra of the vortex-ABS broaden heavily. 
We can also understand the suppression at the Born limit in fig.~\ref{FIG:LDOS-peak}-(d) from this point of view. For this parameter, the minimum excitation energy is finite but very small as in fig.~\ref{FIG:EnergyGap-size}. The narrow gap is insufficient to inhibit resonance between inner and outer states, and as a result the peak broadens even at the Born limit.

When $\Delta_{\text{b}}$ is not so small and there exists a sufficient large gap on the Fermi level in the bulk, quasiparticles with energy lower than the gap edge predominate the vortex-ABS. Considering the impurity scattering between the vortex-ABS only\cite{Kato2000-JPSJ69-3378,Hayashi2005-JLowTempPhys139-79}, we can see that the impurity effects on the vortex-ABS strongly depend on the type of vortex. 
%
We indicate $\delta_{\rm c}$ by the arrows in fig.~\ref{FIG:LDOS-peak} (a)-(d).
We can see in fig.~\ref{FIG:LDOS-peak} (a)-(c) that $\delta_{\rm c}$ moderately well matches the crossover phase-shift from the regime (with smaller $\delta_0$) where the impurity effects depend strongly on the type of vortices to the regime (with larger $\delta_0$) where the vortex-ABS on both types of vortices are heavily suppressed. This result is consistent with our argument in the above. 

\begin{acknowledgements}
  This work was supported by JSPS KAKENHI Grant Number 23244070.
  We thank J.~A.~Sauls for fruitful discussions.
\end{acknowledgements}


\begin{thebibliography}{99}
\bibitem{Maeno1994-Nature372-532}
  Y. Maeno, H. Hashimoto, K. Yoshida, S. Nishizaki, T. Fujita, J. G. Bednorz, F. Lichtenberg, \textit{Nature (London)} \textbf{372}, 532 (1994).
\bibitem{Mackenzie2003-RMP75-657}
  A. P. Mackenzie, Y. Maeno, \textit{Rev.\ Mod.\ Phys.} \textbf{75}, 657 (2003).
\bibitem{Sigrist2005-PTPS160-1} M. Sigrist, \textit{Prog.\ Theor.\ Phys.\ Suppl.} \textbf{160}, 1 (2005).
\bibitem{Maeno2012-JPSJ81-011009}
  Y. Maeno, S. Kittaka T. Nomura, S. Yonezawa, K. Ishida, \textit{J. Phys.\ Soc.\ Jpn.} \textbf{81}, 011009 (2012).
%
%
%
\bibitem{Vorontsov2003-PRB68-064508}
  A. B. Vorontsov, J. A. Sauls, \textit{Phys.\ Rev.\ B} \textbf{68}, 064508 (2003).
%
%
\bibitem{Heeb1999-PRB59-7076}
  R. Heeb, D. F. Agterberg, \textit{Phys.\ Rev.\ B} \textbf{59}, 7076 (1999).
%
\bibitem{Tokuyasu:1} T.~A.~Tokuyasu, D.~W.~Hess, J.~A.~Sauls, \textit{Phys. Rev. B} {\bf 41}, 8891 (1990).
%
%
%
\bibitem{Matsumoto:1} M. Matsumoto, R. Heeb, \textit{Phys. Rev. B} {\bf 65}, 014504 (2001).
%
\bibitem{Kato2001-JPSJ70-3368}
  Y. Kato, N. Hayashi, \textit{J. Phys.\ Soc.\ Jpn.} \textbf{70}, 3368 (2001).
%
\bibitem{Matsumoto1999-JPSJ68-724}  M. Matsumoto, M. Sigrist, \textit{Physica B} \textbf{281-282}, 973 (2000).
%
%
\bibitem{Kato2000-JPSJ69-3378}
  Y. Kato, \textit{J. Phys.\ Soc.\ Jpn.} \textbf{69}, 3378 (2000).
%
%
\bibitem{Kato2002-JPSJ71-1721}
  Y. Kato, N. Hayashi, \textit{J. Phys.\ Soc.\ Jpn.} \textbf{71}, 1721 (2002).
%
%
%
%
\bibitem{Hayashi2005-JLowTempPhys139-79}
  N. Hayashi, Y. Kato, M. Sigrist, \textit{J. Low Temp.\ Phys.} \textbf{139}, 79 (2005).
%
%
\bibitem{Caroli1964-PhysLett9-307}
  C. Caroli, P. G. de Gennes, J. Matricon, \textit{Phys.\ Lett.} \textbf{9}, 30 (1964).
\bibitem{Stone1996-PRB54-13222}
  M. Stone, \textit{Phys.\ Rev.\ B} \textbf{54}, 13222 (1996).
%
%
\bibitem{Tanuma2009-PRL102-117003}
  Y. Tanuma, N. Hayashi, Y. Tanaka, A. A. Golubov, \textit{Phys.\ Rev.\ Lett.} \textbf{102}, 117003 (2009).
%
%
\bibitem{Yokoyama2008-PRL100-177002}
  T. Yokoyama, C. Iniotakis, Y. Tanaka, M. Sigrist, \textit{Phys.\ Rev.\ Lett.} \textbf{100}, 177002 (2008).
%
\bibitem{Tanaka2012-JPSJ81-011013}
  Y. Tanaka, M. Sato, N. Nagaosa, \textit{J. Phys.\ Soc.\ Jpn.} \textbf{81}, 011013 (2012).
%
%
%
\bibitem{SchmittRink}S. Schmitt-Rink, K. Miyake, C. M. Varma, \textit{Phys. Rev. Lett.} \textbf{57}, 2527 (1986).

\bibitem{Hirshfeld}P. Hirshfeld, D. Vollhardt, P. W\"olfle, \textit{Solid State Comm.} \textbf{59}, 111 (1986).
%
\bibitem{Kramer1974-ZPhys269-59}
  L. Kramer, W. Pesch, \textit{Z. Phys.} \textbf{269}, 59 (1974).
\bibitem{Hayashi2013-PhysicaC484-69}
  N. Hayashi, Y. Higashi, N. Nakai, H. Suematsu, \textit{Physica C} \textbf{484}, 69 (2013).
\bibitem{Hayashi2013-PhysicaC}
  N. Hayashi, N. Kurosawa, E. Arahata, Y. Kato, Y. Tanuma, Y. Tanaka, A. A. Golubov, \textit{Physica C} \textbf{494}, 131 (2013).
%
%
\bibitem{Hill2004-PRL92-027001}
  R.~W.~Hill, C.~Lupien, M.~Sutherland, E.~Boaknin, D.~G.~Hawthorn, D.~G.~C.~Proust, F.~Ronning, L.~Taillefer, R.~Liang, D.~A.~Bonn, W.~N.~Hardy, \textit{Phys. Rev. Lett.} \textbf{92}, 027001 (2004).
\bibitem{Sauls2009-NewJPhys11-075008}
  J. A. Sauls, M. Eschrig, \textit{New J. Phys.} \textbf{11}, 075008 (2009).
\bibitem{Eilenberger1968-ZPhys214-195}
  G. Eilenberger, \textit{Z. Phys.} \textbf{214}, 195, (1968).
\bibitem{Thuneberg1984-PRB29-3913}
  E. V. Thuneberg, J. Kurkij\"arvi, D. Rainer, \textit{Phys.\ Rev.\ B} \textbf{29}, 3913 (1984).

\bibitem{Larkin1965-JETPL2-130} A. I. Larkin, \textit{Zh.\ Eksp.\ Teor.\ Fiz.\ Pis'ma Red.} \textbf{2}, 205 (1965) [\textit{JETP Lett.} \textbf{2}, 130 (1965)].%
\bibitem{Mineev}V. P. Mineev, K. V. Samokhin, \textit{Introduction to Unconventional Superconductivity} (Gordon and Breach Science Publishers, 1999).

\bibitem{Nagato1993-JLowTempPhys93-33}
  Y. Nagato, K. Nagai, J. Hara, \textit{J. Low Temp.\ Phys.} \textbf{93}, 33 (1993).
\bibitem{Schopohl1995-PRB52-490}
  N. Schopohl, K. Maki, \textit{Phys.\ Rev.\ B} \textbf{52}, 490 (1995).
\bibitem{SaulsPrivate} J. A. Sauls, private communication.
\bibitem{Maki1999-EPL45-263}
  K. Maki, E. Puchkaryov, \textit{Europhys.\ Lett.} \textbf{45}, 263 (1999).
\end{thebibliography}
\end{document}